\documentclass[twocolumn,prl,showpacs,preprintnumbers,amsmath,amssymb,superscriptaddress]{revtex4}
\usepackage{graphicx}

\begin{document}

\title{On the Physics of Size Selectivity}
\author{Roland Roth} 
\affiliation{Max-Planck-Institut f{\"u}r Metallforschung, Heisenbergstrasse 3,
D-70569 Stuttgart, Germany}
\affiliation{ITAP, Universit{\"a}t Stuttgart, Pfaffenwaldring 57, D-70569 
Stuttgart, Germany} 
\author{Dirk Gillespie}
\affiliation{Department of Molecular Biophysics and Physiology, Rush
University Medical Center, Chicago, Illinois 60612, USA}

\begin{abstract}
We demonstrate that two mechanisms used by biological ion channels to
select particles by size are driven by entropy. With uncharged particles in
an infinite cylinder, we show that a channel that attracts particles is
small-particle selective and that a channel that repels water from the wall is
large-particle selective. Comparing against extensive density-functional
theory calculations of our model, we find that the main physics can be
understood with surprisingly simple bulk models that neglect the confining
geometry of the channel completely. 
\end{abstract}

\maketitle

Ion channels are membrane-spanning proteins that passively transport ions down
their electro-chemical gradients. Channels can open and close their pore upon
stimulation, a process called gating. In addition, some channel types can
preferentially select the ion species they conduct. These properties make
channels responsible for a large number of physiological phenomena including 
propagating action potentials along neurons and initiating muscle contraction.
Experimentally, channels can be studied either in vivo or in well controlled, 
but non-physiological conditions one channel at a time \cite{hille}.

Experimentally, eleven \textquotedblleft selectivity
sequences\textquotedblright\  have been found that rank the relative preference
of a channel for conducting alkali metal ions through cation channels
\cite{hille}. In this paper we focus on the two extreme selectivity sequences
where the channel prefers to conduct small ions or large ions. Examples of 
small-alkali metal cation selective channels include the L-type calcium
channel, the ryanodine receptor (RyR), and the neuronal sodium channel. 
Examples of large-ion selective channels include voltage-gated potassium 
channels, gramicidin A (gA), and the nicotinic acetyl choline receptor (nAchR).

All channel pores are several nanometers in length and have radii ranging from
approximately 2 \AA \ for gA and potassium channels to $>5$ \AA \ for porin 
channels and nAchR. Although its dimensions are small and similar
to the size of ions, equilibrium bulk models for selectivity 
\cite{nce,spmca,lis} that neglect the confining due to the pore can help to
{\em understand} experiments \cite{nce} and reproduces results from Monte
Carlo simulations qualitatively \cite{boda3}. These and other 
studies of selectivity \cite{boda4, pnpdft, porin, ryr} found that
highly-charged channels such as the sodium, L-type calcium, and RyR channels
are small-ion selective because smaller ions balance the charge on the
protein in a smaller volume, a mechanism named charge/space competition. For
large-ion selectivity there are several theories, some based the dehydration
energy of ions \cite{hille, kcsa} and others based on the repulsion of water
from hydrophobic regions of the channel, without explicit ion hydration
\cite{lis}.

In the present Letter we wish to extend the understanding of the physical
mechanism that underlies size selectivity by highlighting entropy as
driving force in both selectivity mechanisms. To do so we choose a model 
with the smallest number of parameters that allows one to study the phenomenon
of size selectivity. This model consists of a mixture of uncharged hard
spheres with different radii \cite{radii}. This model obviously neglects
electrostatics as well as effects caused by hydration shells. In our model the
hard-sphere ``cations'' are attracted into and the hard-sphere ``anions'' are
repelled from the selectivity filter through an {\em effective} potential, 
which parameterizes the long-ranged contribution such as the Donnan potential.
We denote its amplitude by $U_{attr}>0$. We have verified that, due to the 
mentioned repulsion, the presence of anions in the system has a very small
effect on the results. For the sake of simplicity we therefore include only
the solvent and hard-sphere ``cations'' in our model. For studying large-ion
selective channels such as the gA channel we introduce an effective repulsion
$V_{rep}>0$, which models the hydrophobic repulsion of water from the channel. 

Here we focus on the importance of the entropy of both the ions and the
solvent, which is often neglected models of biological systems. Our approach
is in line with the findings of theoretical studies of biological problems in
different areas such as protein folding \cite{Harano04}. If the solvent is
modeled as a fluid of particles with a size comparable to that of the other
species, it provides a crowded environment for the ions, which in turn have to
compete for free space even at low ion concentrations. $U_{attr}$ and
$V_{rep}$, the amplitudes of the ion attraction and the water repulsion, 
respectively, are simply parameters in our model. Since we wish to understand
the {\em phenomenon} of size selectivity among equally charged ions, a
detailed account of electrostatic and hydrophobic interactions will modify the
selectivity of the channel quantitatively, but not the role of entropic forces 
described here.

In order to test if such a simple system selects particles by size, we start by
considering two compartments. The bath, denoted compartment $I$,
contains ions at given concentrations of the order of 100mM dissolved in a
crowded solvent of 55.5 M, the density of pure water under normal
conditions. A second compartment, denoted compartment $II$, models either the
selectivity filter of the channel that attracts ions by the action of
$U_{attr}$ or a hydrophobic channel that repels water by the action of
$V_{rep}$. Inhomogeneities caused by the confining geometry of the protein are
ignored at this stage, but are included later in our density functional
theory (DFT) calculations. 

By allowing equilibrium between the compartments, the concentration of all
components in the filter adjust so that the grand potential of the system is
minimized. This is described by the equality of the chemical potential
$\mu_i^\alpha=\partial f(\{\rho_j^\alpha\})/\partial \rho_i$ of species $i$
between compartments $\alpha=I,II$, where $\rho_i^\alpha$ is the corresponding
number density of this component. For the free energy density of the mixture
$f$ we employ the expression corresponding to the Mansoori Carnahan Starling
and Leland (MCSL) \cite{MCSL} equation of state. 
For the ion components we obtain 
\begin{equation} \label{mu1}
\mu_i^{I}(\{\rho_j^{I}\}) = \mu_i^{II}(\{\rho_j^{II}\})+U_{attr},
\end{equation}
and for the water component 
\begin{equation} \label{mu2}
\mu_{H_2O}^{I}(\{\rho_j^{I}\}) = \mu_{H_2O}^{II}(\{\rho_j^{II}\})-V_{rep}.
\end{equation}
$U_{attr}$ and $V_{rep}$ are the electrostatic and hydrophobic parts of the
chemical potential. 
Equations~(\ref{mu1}) and (\ref{mu2}) are coupled, highly non-linear equations
that determine the densities in the filter compartment $\rho_i^{II}$ for given
densities in the bath, $\rho_i^{I}$, and parameters $U_{attr}$ and $V_{rep}$. 
This bulk model allows us to study the physics of size selectivity assuming
that inhomogeneities in the pore are unimportant. In order to
quantify which component is preferentially absorbed in the pore, we
define the absorbance of component $i$, $\xi_i\equiv \rho_i^{II} /
\rho_i^{I}$, which compares the density of component $i$ in the filter to that
in the bath. The selectivity $S_{i,j}$ of the filter is defined by $S_{i,j}
\equiv \xi_i / \xi_j$. 

The assumption that the inhomogeneities are not important can be tested
by taking the confining geometry of the pore into account within
the framework of DFT. Here we employ a recently improved version of Rosenfeld's
fundamental measure theory which is based on the accurate MCSL equation of 
state \cite{FMT}. In addition to the inputs to our bulk model, the DFT
approach also requires a model for the protein that forms the pore of the
selectivity filter. The simplest way to incorporate the effect of the protein
in the DFT approach would be to define external potentials 
$V_i^{ext}({\bf r})$ that confine particles of all components inside the 
channel. Unfortunately, the actual form of $V_i^{ext}({\bf r})$ is unknown
because of the unknown structure and mechanical properties of the protein. In
our study we choose to model the protein as a hard-sphere fluid which is
restricted by a hard-wall potential to a region {\em outside} the pore. If the
``protein fluid'' is sufficiently dense, spheres of other components will find
it difficult to enter the protein. In this way the protein fluid acts as a
rough wall that can be penetrated by particles of other components, which
is more appropriate than a smooth, hard wall. We model the pore as an
infinitely long cylinder, which simplifies the calculations because the 
density profiles $\rho_i({\bf r})=\rho_i(r)$ of all components, labeled by $i$,
depend only on the radial distance $r$.  

By minimizing the density functional of a 
$N$-component mixture $\Omega[\{\rho_i({\bf r})\}]$
we obtain the inhomogeneous equilibrium density profiles $\rho_i(r)$, as well
as the grand potential $\Omega$ of the system. From these we derive all
quantities of interest. As mentioned above, we consider an external potential
acting only on the protein component, which in turn acts on the other
components in the pore so that they develop inhomogeneous structures. If we
take the limit of the protein fluid density going to zero, no
external potential acts on the system and we recover our bulk
approach. In addition to the parameters entering our bulk approach, two new
parameters are required to describe the system fully, namely the protein
fluid packing fraction $\eta_p$ and the pore radius $R_{pore}$ which defines
the region $r<R_{pore}$ from which the protein fluid is expelled.
Here we chose $\eta_p=0.45$ and the diameter of the particles
that constitute the protein to be 2.45\AA.

In our DFT calculations we obtain the absorbance of species $i$ inside the
pore from its density profile $\rho_i(r)$ by
\begin{equation}
\xi_i = \frac{2}{\rho_i^I R_{pore}^2} \int_0^{R_{pore}} dr r \rho_i(r).
\end{equation}
If the inhomogeneities in $\rho_i(r)$ are small this definition
yields numbers for the absorbance and the selectivity very close to those
predicted by the bulk approach. 

\begin{figure}
\includegraphics[width=0.8\linewidth,clip]{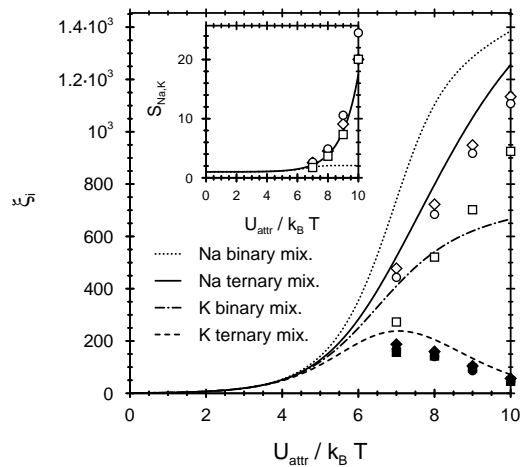}
\caption{\label{fig:sis} The absorbance $\xi_i$ of $Na$ and $K$ in an 
attractive filter as function of $U_{attr}$. We compare bulk results 
(lines) for a binary mixture to those for a ternary mixture. In the
ternary mixture of water, $Na$, and $K$ the competition for space leads to a 
selectivity of the smaller component, $S_{Na,K}\gg 1$, as is shown by the solid
line in the inset. The symbols, which are explained in the text, denote
DFT results for the ternary case.}
\end{figure}

We start by reporting results of our bulk approach which highlight the
role of competition between ion species for free volume in the filter. 
We consider a bath consisting of a binary mixture of 100mM $Na$ {\em or} 100mM
$K$ in water and attract ions into the filter by the potential  $U_{attr}$, in
the range from 0 to 10 $k_B T$. As a result of the attraction, the
concentration of $Na$ or $K$ in the filter increases. We quantify the effect
of the attractive potential by the absorbance $\xi_i$, $i=Na,K$, and show its
dependence on $U_{attr}$ in Fig.~\ref{fig:sis}. The dotted line in
Fig.~\ref{fig:sis} shows the result for the binary mixture of $Na$ and water,
obtained by the bulk approach, and the dashed-dotted line shows the
corresponding quantity for the binary mixture of $K$ and water. Comparing
these results, we observe a similar increase in the filter compartment for
both species, especially at small values of $U_{attr}$, because in that regime
the particles still find free space between the solvent particles. At higher
values of $U_{attr}$, however, $Na$ must squeeze out less water from the
filter than $K$ to follow the attraction, because $\sigma_{Na}<\sigma_K$. The 
result is that $\xi_{Na}>\xi_K$ and hence the selectivity 
$S_{Na,K} = \xi_{Na}/ \xi_K >1$, as is shown in the inset of
Fig.~\ref{fig:sis} (dotted line). An attractive channel favors smaller
species. In this case there is no direct competition between species of
different size and the selectivity $S_{Na,K}$ remains
small, leveling off at 2.2 for large values of $U_{attr}$.

\begin{figure}
\includegraphics[width=0.85\linewidth,clip]{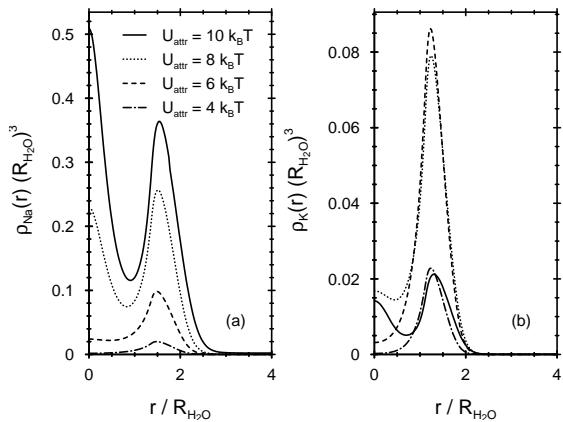}
\caption{\label{fig:prof} Typical density distributions of $Na$ (a) and $K$ (b)
inside an attractive, cylindrical filter as obtained from DFT for various
values of $U_{attr}$. $r$ denotes the radial coordinate. While the density of
$Na$ (a) increases monotonically with $U_{attr}$, the density of $K$ (b) shows
a non-monotonic behavior. For the radius of water we use
$R_{H_2O}=\sigma_{H_2O}/2 = 1.4$\AA \ \cite{radii}.}
\end{figure}

The situation becomes much more interesting and richer if we consider a
ternary mixture of water, 100mM $Na$, {\em and} 100mM $K$ in the bath.
Both ions species feel the attraction  and are in direct competition for the
free volume in the filter. The result of this competition, within the bulk
approach, can be seen 
in Fig.~\ref{fig:sis} (full and dashed lines). For weak attractions the
situation is similar to the previous system, i.e. the ions fill the free
volume between solvent particles and the actual size of the ions is 
unimportant. At stronger attractions, however, when the ions no longer find
free volume, they have to squeeze water out of the filter to
follow the attraction. When the water concentration in the filter is
significantly reduced, direct competition between the ion species sets in and
it becomes energetically and entropically more favorable for the system to
allow the smaller ion species to fill the volume by squeezing out the larger
ions. The $Na$ concentration in the filter continues to increase while the $K$
concentration starts to decrease. If the attraction is sufficiently strong,
this effect can be dramatic and results in a large selectivity $S_{Na,K}\gg 1$,
as we show in the inset of Fig.~\ref{fig:sis} (full line). It is important to
realize that this physical mechanism always prefers the smallest ion species
and prefers $Li$ over $Na$ if we were to start with a bath 
consisting of water, $Li$, and $Na$.  

We test the predictions of our bulk approach by extensive DFT calculations
of our model by changing the pore radius $R_{pore}$ from 1 to
5\AA. Some DFT results for $R_{pore}$=2.1 \AA \ (square), 3.5 \AA \ (circle) 
and 5 \AA \ (diamond) are shown in Fig.~\ref{fig:sis} and its inset. Full 
and open symbols denote results for $K$ and $Na$, respectively. We find that
if the pore is sufficiently wide, the DFT results for the selectivity are in 
good agreement with our bulk predictions, although the absorbances obtained
from DFT deviate slightly from the corresponding bulk values. This agreement
indicates that in this case the inhomogeneities caused by the confinement are
moderate. Most channels whose radii are known have radii $\geq$3
\AA \ \cite{hille, ryr,porin}. If, however, the pore radius becomes smaller
($\lesssim2.5$ \AA) and nearly equal to the particle radii, the confinement
becomes increasingly important and results in deviations between DFT and bulk
results. In this regime, we observe nonlinear absorbance behavior similar to 
that described by Goulding et al. \cite{Goulding01}. In narrow attractive
channels the small ion selectivity can be increased dramatically. Channels
known to have radii this small include gA \cite{hille} and the KcsA potassium
channel \cite{kcsa}.

A set of density profiles of the $Na$ and $K$ species for various values of 
$U_{attr}$ is shown in Fig.~\ref{fig:prof}. The pore radius is 
$R_{pore}=3.5$\AA\ and the ion concentrations in the bath are
$\rho_{Na}=\rho_K=100$mM. These profiles demonstrate the monotonic increase of
$Na$ and the non-monotonic behavior of $K$ in the pore as $U_{attr}$ is
increased. We find that the structure in the density distribution of the 
ions follow closely the densities predicted by the bulk approach.

Now we turn to a different type of channel that is characterized by
hydrophobic protein walls, which we take into account by the effective  
water repulsion $V_{rep}$. Weakly-charged channels such as gA and nAchR seem
to have these properties \cite[and references therein]{lis}. The little charge
in these channels is sufficient to repel anions from the channel but not to
distinguish between cations of different size. In order to keep our
model as simple as possible, we neglect the anions and set
$U_{attr}\equiv 0$, however, we have verified that an extended model that
includes a weak cation attraction and hard-sphere ``anions''
which are repelled from the channel predict equivalent results. Following
Ref.~\cite{lis} we consider $V_{rep}\leq3 k_{B}T$. In the present study we
compare 100mM of both $Na$ and $Cs$, which have a more pronounced size
difference than $Na$ and $K$ \cite{radii}, considered earlier.  

\begin{figure}
\includegraphics[width=0.8\linewidth,clip]{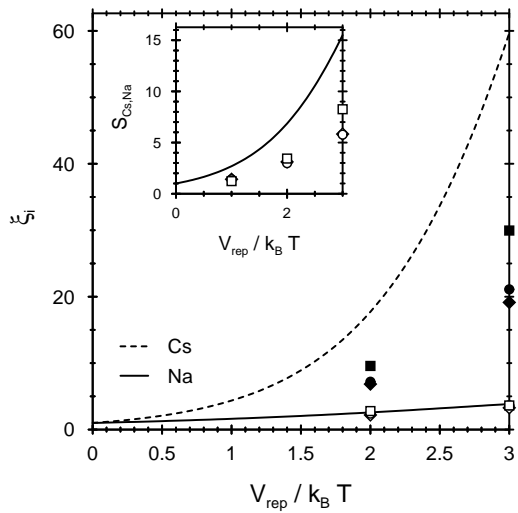}
\caption{\label{fig:lis} The absorbance $\xi_i$ of $Cs$ and $Na$ in a 
hydrophobic filter as function of $V_{rep}$. With increasing value of
$V_{rep}$ both ion densities increase, however, the density of the 
larger component increases faster. The selectivity of $Cs$ over $Na$
is shown in the inset. Lines denote bulk results, while
symbols, which are explained in the text, those obtained by DFT.}
\end{figure}

In Fig.~\ref{fig:lis} we show the absorbance of $Na$ (full line) and $Cs$
(dashed line) in a hydrophobic pore as function of $V_{rep}$ for
a ternary mixture with $\rho_{Na}=\rho_{Cs}=100$mM, predicted by the bulk
theory (lines) and the corresponding selectivity $S_{Cs,Na}$ in the inset. 
While the densities of both species increase monotonically as the water 
repulsion increases, the density of the larger component, $Cs$, increases
faster than that of the smaller component, $Na$. Hence the channel is large ion
selective.

Again we test the prediction of the bulk approach by DFT calculations. In order
to account for the hydrophobic interaction between the protein and water we
introduce an external potential acting on the water. $V^{ext}_{H_2O}(r)$
repels water only from a 1.4 \AA\ neighborhood of the protein. This is a clear
difference between the DFT and the bulk approach, which by construction 
cannot take surface effects into account. This difference is reflected by
lower $Cs$ densities in the pore obtained from DFT (symbols) as
compared to the bulk theory (lines) in Fig.~\ref{fig:lis} and the smaller
selectivity in the inset of Fig.~\ref{fig:lis}. Here the squares, circles and 
diamonds denote DFT results for $R_{pore}=2.8$, 3.5 and 5 \AA , respectively.
This difference gets smaller as the radius of the pore is reduced. The overall
agreement agreement is still qualitatively good. 

Our extensive DFT study confirms the validity of the bulk approach for both 
kinds of channels also for different choices for $\eta_p$ and the ion 
concentrations. The confining geometry of the pore becomes important only
if the radius of the pore is comparable to the radii of particles inside the
pore. Our findings also show clearly that the entropy of the mixture of
particles is enough to give size selectivity and hence has to be taken into
account properly. A very important, yet often neglected, contribution to this
entropy stems from the fact that water is a dense fluid and leaves little
space to the ions. 

RR wishes to thank R.S. Eisenberg for his hospitality and stimulating
discussion during his stay in Chicago, and R. Evans for helpful comments on
the manuscript. DG is grateful for the support of the NIH grant GM 067241 
(Principal Investigator Bob Eisenberg).


\begin{thebibliography}{99} 
\bibitem{hille} B. Hille, {\em Ion Channels of Exciteable Membranes},
(Sinauer Asc., Sunderland, 2001).
\bibitem {nce} W. Nonner, L. Catacuzzeno, and B. Eisenberg, Biophys. J. 
{\bf 79} 1976 (2000). 
\bibitem {spmca} W. Nonner, D. Gillespie, D. Henderson, and B. Eisenberg,
J. Phys. Chem. B. {\bf 105}, 6427 (2001).
\bibitem {lis} D. Gillespie, W. Nonner, D. Henderson, and R.S. Eisenberg,
Phys. Chem. Chem. Phys. {\bf 4}, 4763 (2002).
\bibitem {boda3} see, e.g. D. Boda, D. Henderson, D.D. Busath, and S. 
Soko\l owski, J. Phys. Chem. B. {\bf 104}, 8903 (2000). 
\bibitem {boda4} D. Boda, D.D. Busath, B. Eisenberg, D. Henderson, and W.
Nonner, Phys. Chem. Chem. Phys. {\bf 4}, 5154 (2002).
\bibitem {pnpdft} D. Gillespie, W. Nonner, and R.S. Eisenberg, J. Phys.:
Condens. Matter. {\bf 14}, 12129 (2002). 
\bibitem {porin} H. Miedema, A. Meter-Arkema, J. Wierenga, J. Tang, B.
Eisenberg, W. Nonner, H. Hektor, D. Gillespie, and W. Meijberg,
Biophys. J. {\bf 87},3137 (2004).
\bibitem {ryr} D. Gillespie, L. Xu, Y. Wang, and G. Meissner, J. Phys. Chem. B 
{\bf 109}, 1598 (2005).
\bibitem {kcsa} D.A. Doyle, J. Morais Cabral, R.A. Pfuetzner, A. Kuo, J.M.
Gulbis, S.L. Cohen, B.T. Chait, and R. MacKinnon, Science. {\bf 280},69
(1998).
\bibitem{radii} We take the hard-sphere diameters as the unhydrated crystal
diameter from R.D. Shannon and C.T. Prewitt, Acta Crystallogr. B {\bf 25}, 925
(1969) to be $\sigma_{Na}=2.0$\AA, $\sigma_{K}=2.7$\AA, $\sigma_{Cs}=3.4$\AA,
and $\sigma_{H_2O}=2.8$\AA.
\bibitem{Harano04} Y. Harano and M. Kinoshita, Chem. Phys. Lett. {\bf 399},
342 (2004).
\bibitem{MCSL} G.A. Mansoori, N.F. Carnahan, K.E. Starling, and T.W.
Leland Jr., J. Chem. Phys. {\bf 54}, 1523 (1971).
\bibitem{FMT} Y. Rosenfeld, Phys. Rev. Lett. {\bf 63}, 980 (1989).  R. Roth, 
R. Evans, A. Lang, and G. Kahl, J. Phys.:Condens. Matter \textbf{14}, 12063 
(2002). Y.-X. Yu and J. Wu, J. Chem. Phys. \textbf{117}, 10156 (2002).
\bibitem{Goulding01} D. Goulding, J.-P. Hansen, and S. Melchionna,
 Phys. Rev. Lett {\bf 85}, 1132 (2000).
\end{thebibliography}
\end{document}